\documentclass[11pt, fceqn]{article}
 \usepackage{amsfonts}
 \usepackage{flafter}
 \usepackage{soul, color, xcolor}
 \usepackage{amssymb,graphics,graphicx,subfigure,caption2,rotating}
 \usepackage{amsmath, latexsym}
 \usepackage[amsmath,thmmarks]{ntheorem}
 \usepackage[top=3.5cm]{geometry}
 \usepackage{indentfirst}
 \usepackage{mathrsfs}
 \usepackage{subfigure}
 \usepackage[section]{placeins}
 \newcommand\keywords[1]{Keywords: #1}
 \allowdisplaybreaks[4]
\begin{document}
\date{}
	\title{\textbf{Variational quantum algorithm for the Poisson equation based on the banded Toeplitz systems}}
	\author{Xiaoqi Liu,  Yuedi Qu,  Ming Li\thanks{liming@upc.edu.cn.} and Shu-qian Shen \\
		{\footnotesize{\it College of Science, China University of Petroleum, Qingdao 266580, China}}
	}\maketitle
	\begin{abstract}
For solving the Poisson equation it is usually possible to discretize it into solving the corresponding linear system $Ax=b$. Variational quantum algorithms (VQAs) for the discreted Poisson equation have been studied before. We give a VQA based on the banded Toeplitz systems for solving the Poisson equation with respect to the structural features of matrix $A$. In detail, we decompose the matrix $A$ and $A^2$ into a linear combination of the corresponding banded Toeplitz matrix and sparse matrices with only a few non-zero elements. For the one-dimensional Poisson equation with different boundary conditions and the $d$-dimensional Poisson equation with Dirichlet boundary conditions, the number of  decomposition terms is less than the work in [Phys. Rev. A108, 032418 (2023)]. Based on the decomposition of the matrix, we design quantum circuits that evaluate efficiently the cost function. Additionally, numerical simulation verifies the feasibility of the proposed algorithm. In the end, the VQAs for linear systems of equations and matrix–vector multiplications with $K$-banded Teoplitz matrix $T_n^K$ are given, where $T_n^K\in R^{n\times n}$ and $K\in O({\rm ploy}\log n)$.

\end{abstract}
\maketitle

\section{Introduction}

From the earliest algorithms for factoring large numbers \cite{Shor's algorithm}, unstructured database searching \cite{Grover's algorithm} to algorithms for linear systems of equations \cite{HHL} and quantum machine learning \cite{QML}, quantum computing has demonstrated its tremendous computational advantages. However, realizing some quantum algorithms that require large depth quantum circuits consisting of high precision quantum gates is still difficult in the current era of noisy intermediate-scale quantum (NISQ) devices \cite{NISQ}. Existing quantum devices contain about 100 physical qubits, which is not perfect since the qubits and quantum operations supported by NISQ devices are not error corrected \cite{VQAs}. Even without error correction, noise in shallow circuit implementations can be suppressed by error mitigation \cite{error1,error2}, suggesting that it is feasible to utilize NISQ hardware for quantum computing. Most current NISQ algorithms delegate the classically difficult part of the computation to a quantum computer, while the other part of the computation is performed on a sufficiently powerful classical device. These types of quantum algorithms are known as variational quantum algorithms (VQAs) \cite{VQAs}. The earliest VQAs can be traced back to variational quantum eigensolver(VQE) \cite{VQE} and quantum approximate optimization algorithm (QAOA) \cite{QAOA}. Afterwards, inspired by the existing algorithms, VQAs in various fields such as machine learning and quantum information are proposed.\\

The Poisson equation has important applications in many areas of science and engineering, such as computational fluid dynamics \cite{Fluid_Dynamics} and theory of Markov chains \cite{Control_Techniques,Markov_chains}, and are also important for density functional theory and electronic structure calculations \cite{density}. It is therefore very interesting to explore how the Poisson equation can be solved on NISQ devices. In general, the Poisson equation is discretized using methods such as the finite-difference method \cite{FDM} to obtain a linear system whose solution approximates the Poisson equation. VQAs for solving linear systems on near-term quantum devices have been proposed \cite{VQLS,QALS,HEVQE,VALA,dynamic_ansatz}. Most of these VQAs assume that the coefficient matrix of the linear system can consist of a polynomial number (with respect to the number of qubits) of tensor products of the local operator, such as, $I, \sigma _X,\sigma _Y,\sigma _Z$. For general linear systems, to find a strategy satisfying the above requirements is a nontrivial problem. Based on the structure of the coefficient matrix of the discretized Poisson equation, work has been done to design relevant variational quantum algorithms in a targeted manner. The authors in  \cite{poisson2021} find an explicit tensor product decomposition of coefficient matrix under a set of simple operators $\{I, \sigma_+ =|0\rangle\langle1|, \sigma_- =|1\rangle\langle 0|\}$, with only $(2\log 2n+1)$ items,  where $n$ is the dimension of the coefficient matrix $A$. Inspired by \cite{poisson2021}, the authors in \cite{sigma_basis}, develop a method for decomposing matrices based on the sigma basis. Yuki Sato et al. \cite{poisson_energy} propose a variational quantum algorithm based on minimal potentials to solve the Poisson equation, which decomposes the corresponding observations into linear combinations of the tensor product of Pauli operators and simple observables. In article \cite{poisson2023}, the authors rewrite the cost function as a quantity related to $\sigma _X\otimes A$ and decompose $\sigma_X\otimes A$ into a sum of Hermitian, one-sparse, and self-inverse operators. Then, for the one-dimensional Poisson equation with different boundary conditions and for the $d$-dimensional Poisson equation with Dirichlet boundary conditions, $\sigma _X\otimes A$ is decomposed into a sum of at most $7$ terms and $4d+1$ terms respectively, and $\sigma _X\otimes A^2$ is decomposed into a sum of at most $15$ terms and $ (4d+1)^2-(4d+1)$ terms respectively. And the number of terms in the sparse matrix factorization is independent of the coefficient matrix dimension $n$.\\

In this paper,  we present a different decomposition strategy to reduce the number of the decomposition terms for solving Poisson equation on NISQ devices. We transform the matrix $A$ into the banded Teoplitz matrix and give a VQA for the banded Teoplitz system. Then, for the one-dimensional Poisson equation with different boundary conditions, we decompose the corresponding Hamiltonian into at most $5$ and $6$ terms. Similarly, the corresponding Hamiltonian of the $d$-dimensional Poisson equation with Dirichlet boundary conditions can be decomposed into at most $4d+1$ and $12d^2$. Naturally, the decomposition method and the quantum circuit design remain applicable for linear system with Hermitian and sparse coefficient matrices satisfying $a_{i,i+c}=a_c$ for $c=0, 1, \cdots, n-1$ and $i=0, \cdots, n-1-c$, where $a_{i,i+c}$ denotes the element of matrix $A$.\\

The remainder of the paper is organized as follows. In Sec. 2, we briefly review the definition of the Poisson equation, the discretized Poisson equation, and the definition of the corresponding cost function. In Sec. 3, we expound how to approximate the ground state of the Hamiltonian by using VQA based on the decomposition of $A$ and $A^2$, and numerically illustrate the performance of our algorithm for the one-dimensional Poisson equation with Dirichlet boundary conditions. We generalize the algorithm of the Poisson equation to the banded Teoplitz matrix in Sec. 4. Finally, we present our conclusion and discussion in Sec. 5.

\section{The VQA for the Poisson equation}
The one-dimensional Poisson equation with different boundary conditions \cite{poisson2021,poisson2023} can be written as
\begin{gather*}
-\bigtriangleup \mu (x)=f(x), x\in(0,1),
\end{gather*}
with the unified boundary conditions $\alpha_1{\mu }' (0)-\alpha_2\mu(0)=0$ and $\beta _1{\mu }' (0)-\beta_2\mu(0)=0$, where $\bigtriangleup$ is the Laplace operator; $\alpha_1$, $\alpha_2$, $\beta _1$ and $\beta _2$ are all positive constants; and $f: D\to  R$ is a sufficiently smooth function. The $d$-dimensional Poisson equation with Dirchlet boundary conditions is of the form $-\bigtriangleup \mu (x)=f(x)$, $x\in(0,1)^d$ with the boundary condition $\mu (x)=0$, $x\in\{0,1\}^d$.\\

By the finite-difference method, the discretized one-dimensional Poisson equation with the unified boundary conditions is 
\begin{gather*}
\tilde{A}x=b,
\end{gather*}
where
\begin{gather*}
\tilde{A}=\begin{bmatrix}
  2-c      &  -1     &\cdots   & 0\\
  -1     &  \ddots & \ddots  & \vdots \\
  \vdots & \ddots  &\ddots   &-1 \\
  0      &  \cdots & -1      &2-d
\end{bmatrix}\in R^{n\times n},
\end{gather*}
and $c=\frac{\alpha _1}{\alpha _1+\alpha _2h}$, $d=\frac{\beta  _1}{\beta  _1+\beta  _2h}$, $h=\frac{1}{n+1}$, $n$ comes from evenly dividing $(0,1)$ into $n+1$ parts during the discretization, and $b\in R^n$ is the vector obtained by sampling $f(x)$ on the interior grid points. For simplicity, we assume that $n = 2^N$, where $N$ is a positive integer. Similarly, we can also obtain the coefficient matrix generated by the discretization of the $d$-dimensional Poisson equation with Dirichlet boundary conditions,\\
\begin{gather*}
A^{(d)}=A'\otimes  I\otimes  \dots \otimes I+I\otimes A'\otimes I\otimes \dots \otimes I+\cdots +I\otimes \cdots \otimes I\otimes A',
\end{gather*}
where 
\begin{gather*}
A'=\begin{bmatrix}
  2     &  -1     &\cdots   & 0\\
  -1     &  \ddots & \ddots  & \vdots \\
  \vdots & \ddots  &\ddots   &-1 \\
  0      &  \cdots & -1      &2
\end{bmatrix}\in R^{n\times n},
\end{gather*}
$I\in R^{n\times n}$ and $A^{(d)}\in R^{n^d\times n^d}$.\\

As with most VQAs for solving linear system $Ax=b$, we transform it to find the ground state of a Hamiltonian $H$,
\begin{gather*}
H=A^\dagger (I-|b \rangle \langle b|)A,
\end{gather*}
where $\left | b  \right \rangle \propto b$, $A$ is the coefficient matrix of the linear system. And we assume that there is an efficient unitary operator $U_b$ that can prepare a quantum state $\left | b  \right \rangle$.
It is easy to verify that the solution $\left | x  \right \rangle$ of a linear system is the unique eigenstate corresponding to the minimum eigenvalue $0$ of Hamiltonian $H$.

To obtain $\left | x  \right \rangle$, we define the cost function $E(\theta)=\langle\psi(\theta)|H|\psi(\theta)\rangle$, where $|\psi(\theta)\rangle=U(\theta)|0\rangle$, with unitary gate $U(\theta)=\prod_{i=1}^{d}\otimes_{t=1}^NU(\theta_t^i)W_i$, where $U(\theta_t^i)$ is a product of combinations of single-qubit rotational gates, $\theta=[\theta_1^0,\dots,\theta_N^0,\dots,\theta_1^d,\dots,\theta_N^d]$, $d$ is the depth of the quantum circuit, $\theta_t^i\in [0,2\pi]$ and $W_i$ is composed of nearest-neighbor CNOT gate or CZ gate. The cost function $E(\theta)$ represents the expectation value of $H$ under the state $|\psi(\theta)\rangle$. Then we aim to optimize the parameters $\theta$ to minimize $E(\theta)$, that is,
\begin{align*}
		\underset{\theta}{\min} E(\theta)&=\underset{\theta}{\min}\langle\psi(\theta)|H|\psi(\theta)\rangle\\
                                           &=\underset{\theta}{\min}[\langle\psi(\theta)|A^2|\psi(\theta)\rangle-\left  |\langle b|A|\psi(\theta)\rangle\right | ^2].
\end{align*}

To design a VQA for the Poisson equation, we give the decomposition based on the characterization of the matrix $A$ and $A^2$.

\section{An explicit decomposition of $A$ and $A^2$}
\subsection{One-dimensional Poisson equation}
Before decomposing the matrices $A$ and $A^2$, we introduce a special class of matrices, i.e., Teoplitz matrices \cite{circulant_like_operators,Toe}. The Teoplitz matrix $T_n$ is a matrix of size $n\times n$ whose elements along each diagonal are constant. More clearly,
\begin{gather*} 
T_n=\begin{bmatrix}
  t_0&  t_{-1}& t_{-2}&    \cdots & t_{-(n-1)}\\
  t_1&  t_0&  t_{-1}&    \cdots & t_{-(n-2)}\\
  t_2&  t_1  &  t_0&\ddots    & \vdots \\
  \vdots &  \ddots &  \ddots &   \ddots &t_{-1} \\
  t_{(n-1)} & \cdots &  t_2&  t_1&t_0
\end{bmatrix},
\end{gather*}
where $t_{i,k}=t_{i-k}$ and $T_n$ is decided with $2n-1$ elements, i.e., $\{t_j\} _{j=-(n-1)}^{n-1}$ . Specially, when $t_j=0$, $j=\pm (K+1),\dots,\pm (n-1)$, $T_n$ is a banded Teoplitz matrix $T_n^K$, namely, $2K+1$-sparse diagonal matrix.\\

The circulant matrix is a special form of the Teoplitz matrix. Its form is
\begin{gather*} 
C_n=\begin{bmatrix}
  c_0&  c_{n-1}& c_{n-2}&    \cdots & c_1\\
  c_1&  c_0&  c_{n-1}&    \cdots & c_2\\
  c_2&  c_1  &  c_0&\ddots    & \vdots \\
  \vdots &  \ddots &  \ddots &   \ddots &c_{n-1}\\
  c_{n-1} & \cdots &  c_2&  c_1&c_0
\end{bmatrix}.
\end{gather*} 

When $c_{K+1},c_{K+2},\dots,c_{n-K-1}=0$, $C_n$ is a $2K+1$-sparse matrix $C_n^K$. In this paper, it is necessary to rely on a special kind of circulant matrix, i.e., the unit circulant matrix:
\begin{gather*} 
L=\begin{bmatrix}
  0&  0&  \cdots &0  &1 \\
  1&  0&  &  &0 \\
  \vdots &  \ddots &  \ddots &  &\vdots  \\
  \vdots &  &  1&  0&0  \\
  0&  \cdots &  \cdots &  1&0
\end{bmatrix},
\end{gather*} 
\begin{gather*} 
R=\begin{bmatrix}
  0&  1&  \cdots &0  &0 \\
  0&  0&1  &  &0 \\
  \vdots &   &  \ddots &\ddots  &\vdots  \\
  0 &  &  &  0&1  \\
  1&  0 &  \cdots &  0&0
\end{bmatrix}.
\end{gather*} 

Note that $L^{\dagger}=R=L^{-1}$ and $L^{n-l}=L^{-l}$, and
\begin{gather*}
C_n^K=\sum_{l=0}^{K} c_lL^l + \sum_{\imath =1}^{K} c_{n-\imath }R^\imath 
=\sum_{\substack{l=0\\l\ne K+1,\dots,n-K+1}}^{N-1} c_lL^l=\sum_{l=-K}^{K} c_lL^l,
\end{gather*}
where $c_{-l}=c_{n-l}$.
Any circulant matrix $C_n$ can be decomposed as 
\begin{gather*}
C_n=F_n^{-1}{\rm diag}(F_nc)F_n,
\end{gather*}
where the matrix $F_n$ is a Fourier transform of order $n$ and the vector $c=[c_0, c_1,\dots,c_{n-1}]$ is the first column of the circulant matrix $C_n$.
In particular, for the unit circulant matrix $L$, one has $L=F_n^{-1}DF_n$, where
\begin{gather*}
D={\rm diag}\{\omega _{n}^{0} ,\omega _{n}^{1},\cdots ,\omega _{n}^{n-1}\},
\end{gather*}
$\omega _{n}^{i}$ is the $n$th unit root, and can also be written as
\begin{gather*}
D=\begin{pmatrix}
  1&0 \\
  0&\omega _{n}^{n/2}
\end{pmatrix}\otimes 
\dots\otimes 
\begin{pmatrix}
  1&0 \\
  0&\omega _{n}^{2}
\end{pmatrix}
\otimes 
\begin{pmatrix}
  1&0 \\
  0&\omega _{n}^{0}
\end{pmatrix}
=\otimes _{j=0}^{N-1} P(\theta _j),
\end{gather*}
where
\begin{gather*}
P(\theta)=\begin{bmatrix}
  1& 0\\
  0&{\rm e}^{{\rm i}\theta}
\end{bmatrix},
\end{gather*}
$\theta _j=\frac{2\pi}{n} 2^j$.
Thus, computing the expectation of a Toeplitz matrix $T_n$ can be obtained by embedding it in a $2n\times2n$ circulant matrix $C_{2n}$, i.e.,
\begin{gather*}
C_{2n}=\begin{bmatrix}
  T_n&B_n   \\
  B_n&T_n
\end{bmatrix},
\end{gather*}
where
\begin{gather*}
B_n=\begin{bmatrix}
  0&  t_{n-1}&\cdots   & t_{2} & t_{1} \\
  t_{-(n-1)}& 0&  t_{n-1}& \cdots & t_{2}\\
  \vdots &  t_{n-1} &0  &\ddots   & \vdots\\
  t_{-2}&  &  \ddots & \ddots  &t_{n-1} \\
  t_{-1}&  t_{-2}&  \cdots& t_{-(n-1)} &0
\end{bmatrix}.
\end{gather*}

Then, the expectation value of the circulant matrix $C_{n}^K$ can be expressed as
\begin{align*}
\langle \psi(\theta)|C_{n}^K|\psi(\theta)\rangle&=\sum_{l=-K}^{K} c_l\langle \psi(\theta)|L^l|\psi(\theta)\rangle\\
&=\sum_{l=-K}^{K} c_l\langle \psi(\theta)|F_n^{-1}D^lF_n|\psi(\theta)\rangle,
\end{align*}
where 
\begin{gather*}
D^l={\rm diag}\{(\omega _{n}^{0})^l ,(\omega _{n}^{1})^l,\cdots ,(\omega _{n}^{n-1})^l\}.
\end{gather*}

We will show the decompositions of matrices $A$ and $A^2$ according to the special structures. The main idea is to transform them into banded Teoplitz matrices for solving, and this process produces sparse matrices with only a small number of  non-zero elements, for which we will make use of the extended Bell measurements method \cite{EBM}.\\ 

In particular, when the one-dimensional Poisson equation has Dirichlet bound boundary conditions, the matrix of the corresponding linear system is $A'$ which is a general tridiagonal Toeplitz matrix $T_n^1$.

Thus, $\left  |\langle b|A'|\psi(\theta)\rangle\right | ^2$ can be obtained by computing $\left  |\langle 0,b|C_{A'}|0,\psi(\theta)\rangle\right | ^2$:
\begin{equation}
\begin{split}
&\left  |\langle b|A'|\psi(\theta)\rangle\right | ^2=\left  |\langle b|T_n^1|\psi(\theta)\rangle\right | ^2=\left  |\langle 0,b|C_{T_n^1}|0,\psi(\theta)\rangle\right | ^2\\
=&|-\langle 0,b|F_n^{-1}DF_n|0,\psi(\theta)\rangle - \langle 0,b|F_n^{-1}D^{2n-1}F_n|0,\psi(\theta)\rangle +2 \langle 0,b|0,\psi(\theta)\rangle | ^2,\label{eq1}
\end{split}
\end{equation}
where
\begin{gather*}
C_{A'}=\begin{bmatrix}
  A'&B\{A'\}_n   \\
  B\{A'\}_n&A'
\end{bmatrix},
\end{gather*}
\begin{gather*}
B\{A'\}_n=\begin{bmatrix}
  0&  0   &\cdots   &0  &-1 \\
  0&  0&  0& \ddots    &0 \\
  \vdots & &\ddots   &  &\vdots  \\
  0&   &  &  \ddots &0 \\
 -1&0  &\cdots   &  0&0
\end{bmatrix}.
\end{gather*}

Similarly, ${A'}^2$ can be expressed as
\begin{align*}
{A'}^2&=\begin{bmatrix}
5&  -4&  1&  &  & 0 \\
  -4&  6&  -4&  1&  & \\
  1&  \ddots &  \ddots &  \ddots &\ddots   & \\
  &  \ddots &  \ddots &  \ddots &  \ddots & 1 \\
  &  &  1&  -4&  6& -4 \\
  &  &  &  1&  -4&5
\end{bmatrix}\\
&=
\begin{bmatrix}
  6&  -4&  1&  &  & 0 \\
  -4&  6&  -4&  1&  & \\
  1&  \ddots &  \ddots &  \ddots &\ddots   & \\
  &  \ddots &  \ddots &  \ddots &  \ddots & 1 \\
  &  &  1&  -4&  6& -4 \\
  &  &  &  1&  -4& 6
\end{bmatrix}
-\begin{bmatrix}
 1&  &  &  &  &  \\
 &  0&  &  &  & \\
  &   &  \ddots & &  & \\
  &   &  &  \ddots &   &  \\
  &  &  &  &0 &  \\
  &  &  &  & &1
\end{bmatrix}\\
&=T_n^2-M_1.
\end{align*}

And $\langle\psi(\theta)|{A'}^2|\psi(\theta)\rangle$ can be obtained by calculating
\begin{align*}
 \langle\psi(\theta)|T_n^2-M_1|\psi(\theta)\rangle&= \langle\psi(\theta)|T_n^2|\psi(\theta)\rangle - \langle\psi(\theta)|M_1|\psi(\theta)\rangle,\\
\end{align*}
\begin{equation}
\begin{split}
\langle\psi(\theta)|T_n^2|\psi(\theta)\rangle=&\langle0,\psi(\theta)|C_{T_n^2}|0,\psi(\theta)\rangle\\
=&-4\langle 0,\psi(\theta)|F_n^{-1}DF_n|0,\psi(\theta) \rangle\\
&+\langle 0, \psi(\theta)|F_n^{-1}D^2F_n|0, \psi(\theta) \rangle\\
&-4\langle 0, \psi(\theta)|F_n^{-1}D^{2n-1}F_n|0, \psi(\theta) \rangle\\
&+\langle 0, \psi(\theta)|F_n^{-1}D^{2n-2}F_n| 0,\psi(\theta) \rangle+6\\
=&-4\langle 0,\psi(\theta)|F_n^{-1}DF_n|0,\psi(\theta) \rangle+\langle 0, \psi(\theta)|F_n^{-1}D^2F_n|0, \psi(\theta) \rangle\\
&-4\overline{\langle 0, \psi(\theta)|F_n^{-1}DF_n|0, \psi(\theta) \rangle}+\overline{\langle 0, \psi(\theta)|F_n^{-1}D^2F_n| 0,\psi(\theta) \rangle}+6,\label{eq2}
\end{split}
\end{equation}
where
\begin{gather*}
C_{T_n^2}=\begin{bmatrix}
  T_n^2&B\{T_n^2\}_n   \\
  B\{T_n^2\}_n&T_n^2
\end{bmatrix},
\end{gather*}
\begin{gather*}
B\{T_n^2\}_n=\begin{bmatrix}
  0&  0&  &  &  1& -4 \\
  0&  0&  &  &  0&1 \\
  &  \ddots &  \ddots &  \ddots &\ddots   & \\
  &  \ddots &  \ddots &  \ddots &  \ddots &  \\
  1&  0&  &  &  0& 0 \\
  -4&  1&  &  &  0& 0
\end{bmatrix}.
\end{gather*}

For the sparse matrix $M_1$, due to the specificity (symmetry) of the positions of its non-zero elements, we can construct the associated quantum circuit. We take $n=8$ ($N=3$) for example to illustrate the quantum circuit. The matrix $M_1$ can be rearranged as
\begin{align*}
M_1=&|000\rangle\langle 000|+|111\rangle\langle 111|\\
=&\frac{1}{2} (|000\rangle+|111\rangle)(\langle 000|+\langle 111|)+\frac{1}{2} (|000\rangle-|111\rangle)(\langle 000|-\langle 111|)\\
=&({\rm CNOT}_{2}^{0}{\rm CNOT}_{1}^{0}H^{0}|000\rangle)({\rm CNOT}_{2}^{0}{\rm CNOT}_{1}^{0}H^{0}|000\rangle)^\dagger\\
&+({\rm CNOT}_{2}^{0}{\rm CNOT}_{1}^{0}H^{0}X^{0}|000\rangle)({\rm CNOT}_{2}^{0}{\rm CNOT}_{1}^{0}H^{0}X^{0}|000\rangle)^\dagger,
\end{align*}
where $G_{j}^{i}$ represents the quantum gate $G$ applied at $i$th qubit controlled by the $j$th qubit, such as ${\rm CNOT}_{j}^{i}$ is the CNOT gate whose control and target qubits are $i$th 
and $j$th qubits, respectively.\\

Thus, for $\langle\psi(\theta)|M_1|\psi(\theta)\rangle$, we have 
\begin{align*}
\langle\psi(\theta)|M_1|\psi(\theta)\rangle
=&\langle\psi(\theta)|({\rm CNOT}_{2}^{0}{\rm CNOT}_{1}^{0}H^{0}|000\rangle)({\rm CNOT}_{2}^{0}{\rm CNOT}_{1}^{0}H^{0}|000\rangle)^\dagger\\
&+({\rm CNOT}_{2}^{0}{\rm CNOT}_{1}^{0}H^{0}X^{0}|000\rangle)({\rm CNOT}_{2}^{0}{\rm CNOT}_{1}^{0}H^{0}X^{0}|000\rangle)^\dagger|\psi(\theta)\rangle\\
=&|\langle\psi(\theta)|{\rm CNOT}_{2}^{0}{\rm CNOT}_{1}^{0}H^{0}|000\rangle|^2+|\langle\psi(\theta)|{\rm CNOT}_{2}^{0}{\rm CNOT}_{1}^{0}H^{0}X^{0}|000\rangle|^2.\\
\end{align*}

The quantum circuit of the one-dimensional Poisson equation with Dirichlet bound boundary conditions is shown in Figure 1.


When the Poisson equation has the unified boundary conditions, the matrix of the corresponding linear system is $\tilde{A}$.
Decomposing $\tilde{A}$, we have
\begin{align*}
\tilde{A}
&=\begin{bmatrix}
  2&  -1   &\cdots   &0  &0 \\
  -1&  2&  -1& \ddots    &0 \\
  \vdots & &\ddots   &  &\vdots  \\
  0&   &  &  \ddots &-1 \\
  0&0  &\cdots   &  -1&2
\end{bmatrix}
-c\begin{bmatrix}
  1      & 0     &\cdots   & 0\\
  0   &  \ddots & \ddots  & \vdots \\
  \vdots & \ddots  &\ddots   &0 \\
  0      &  \cdots &0      &0
\end{bmatrix}
-d\begin{bmatrix}
  0     & 0     &\cdots   & 0\\
  0   &  \ddots & \ddots  & \vdots \\
  \vdots & \ddots  &\ddots   &0 \\
  0      &  \cdots &0      &1
\end{bmatrix}\\
&=A'-cM_2-dM_3\\
&=T_n^1-cM_2-dM_3.
\end{align*}

\begin{figure}[!h]
		\centering
        \subfigure[]{\includegraphics[width=0.5\textwidth]{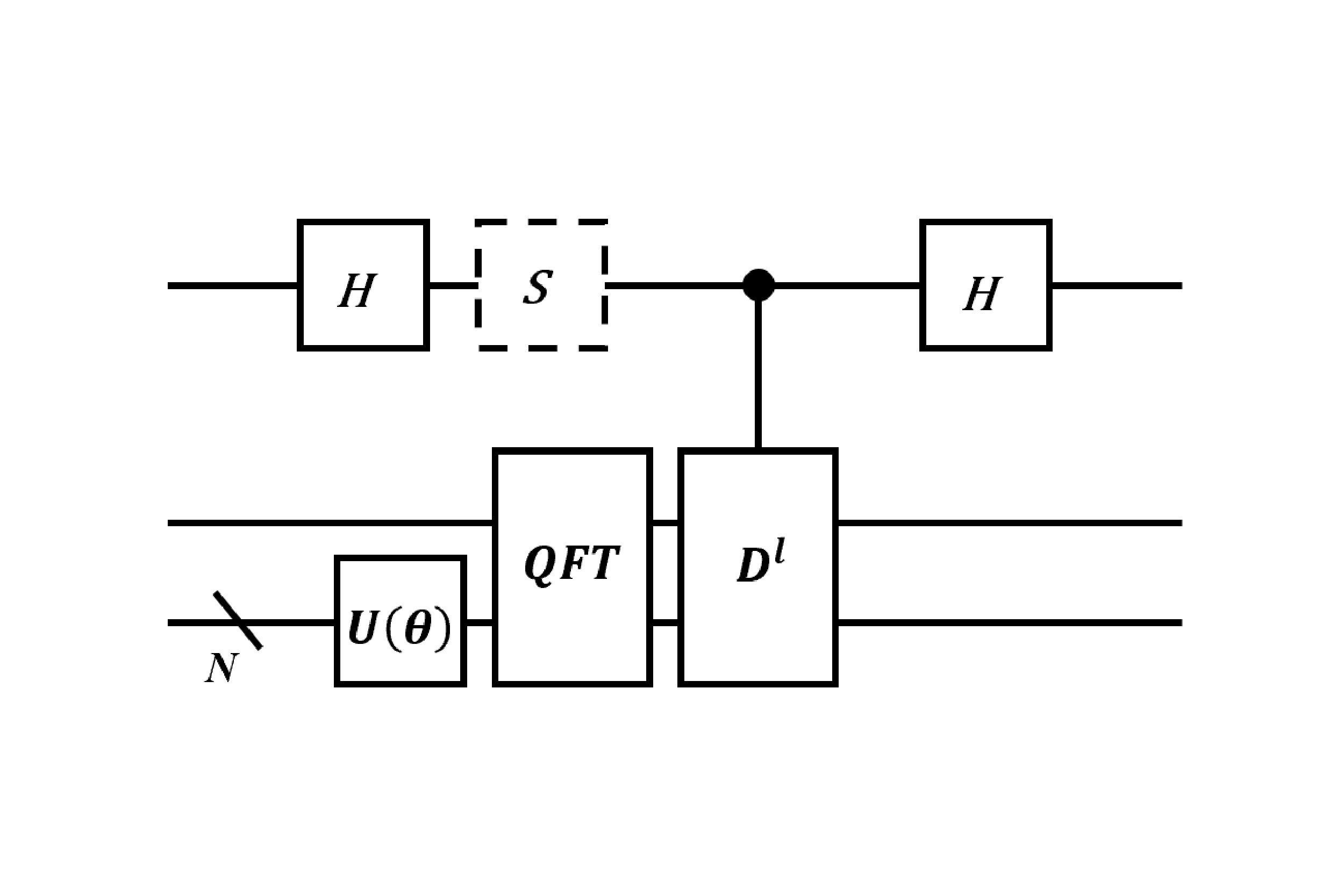}\label{fig1a}}\subfigure[]{\includegraphics[width=0.5\textwidth]{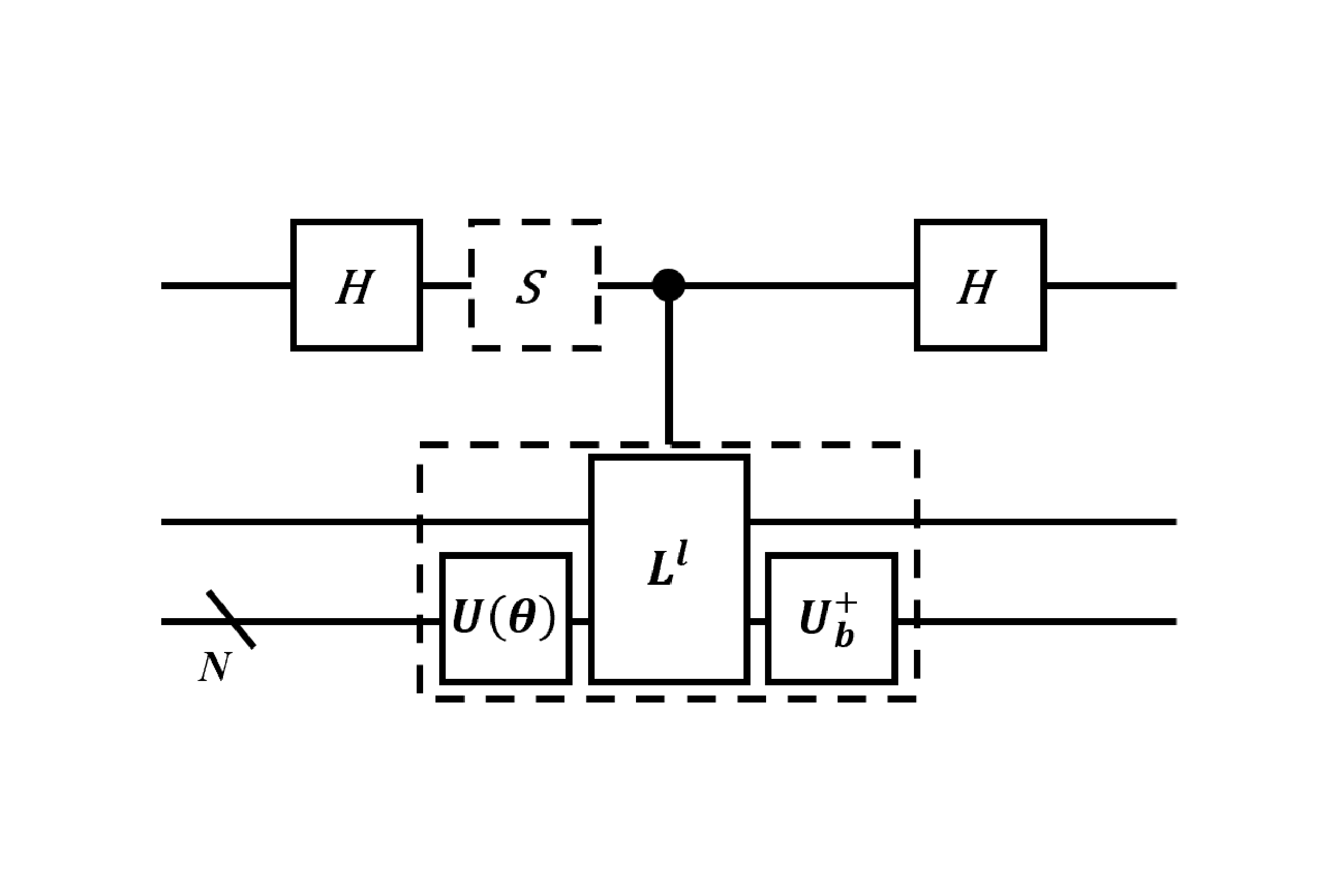}\label{fig1b}}
		\subfigure[]{\includegraphics[width=0.5\textwidth]{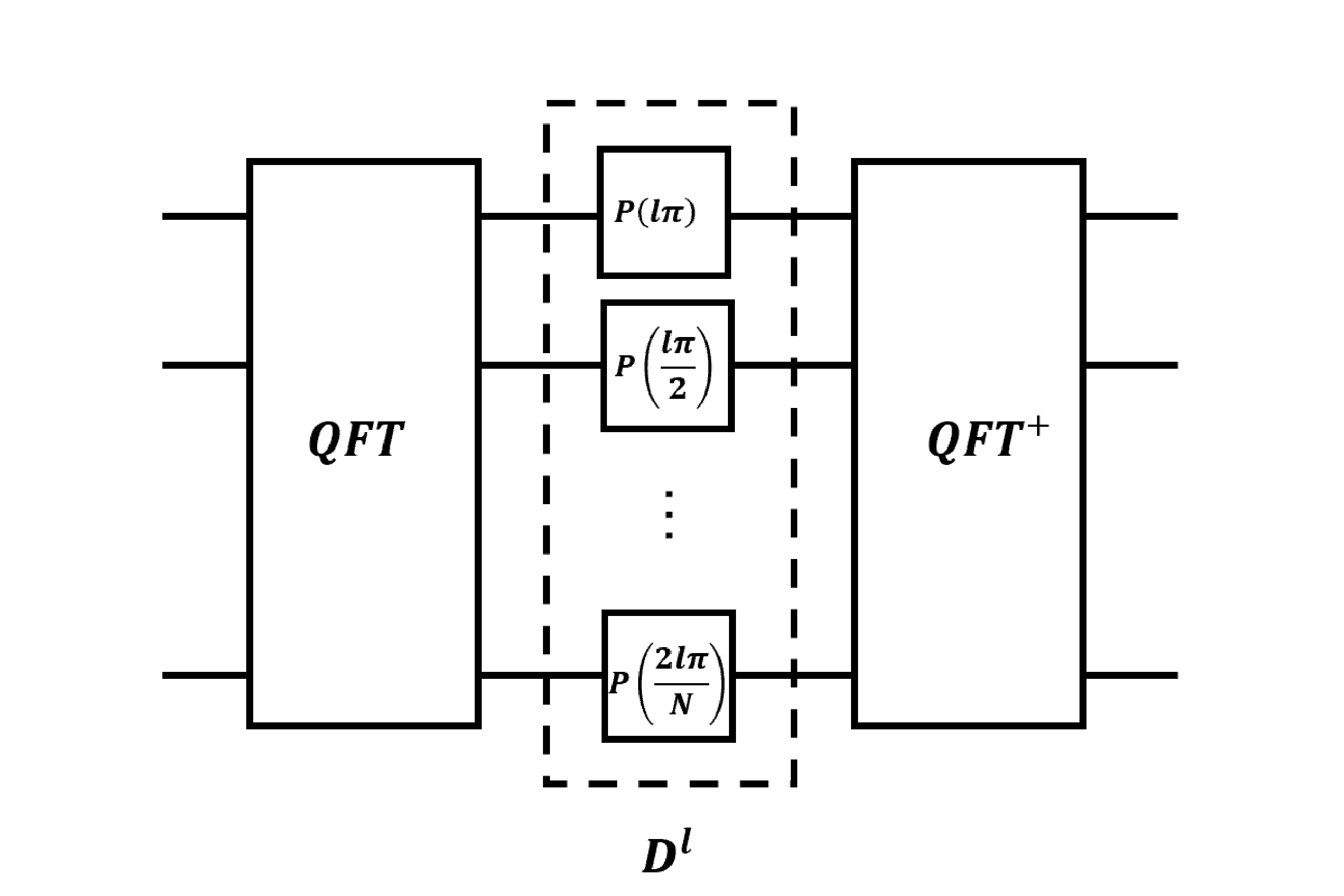}\label{fig1c}}\subfigure[]{\includegraphics[width=0.5\textwidth]{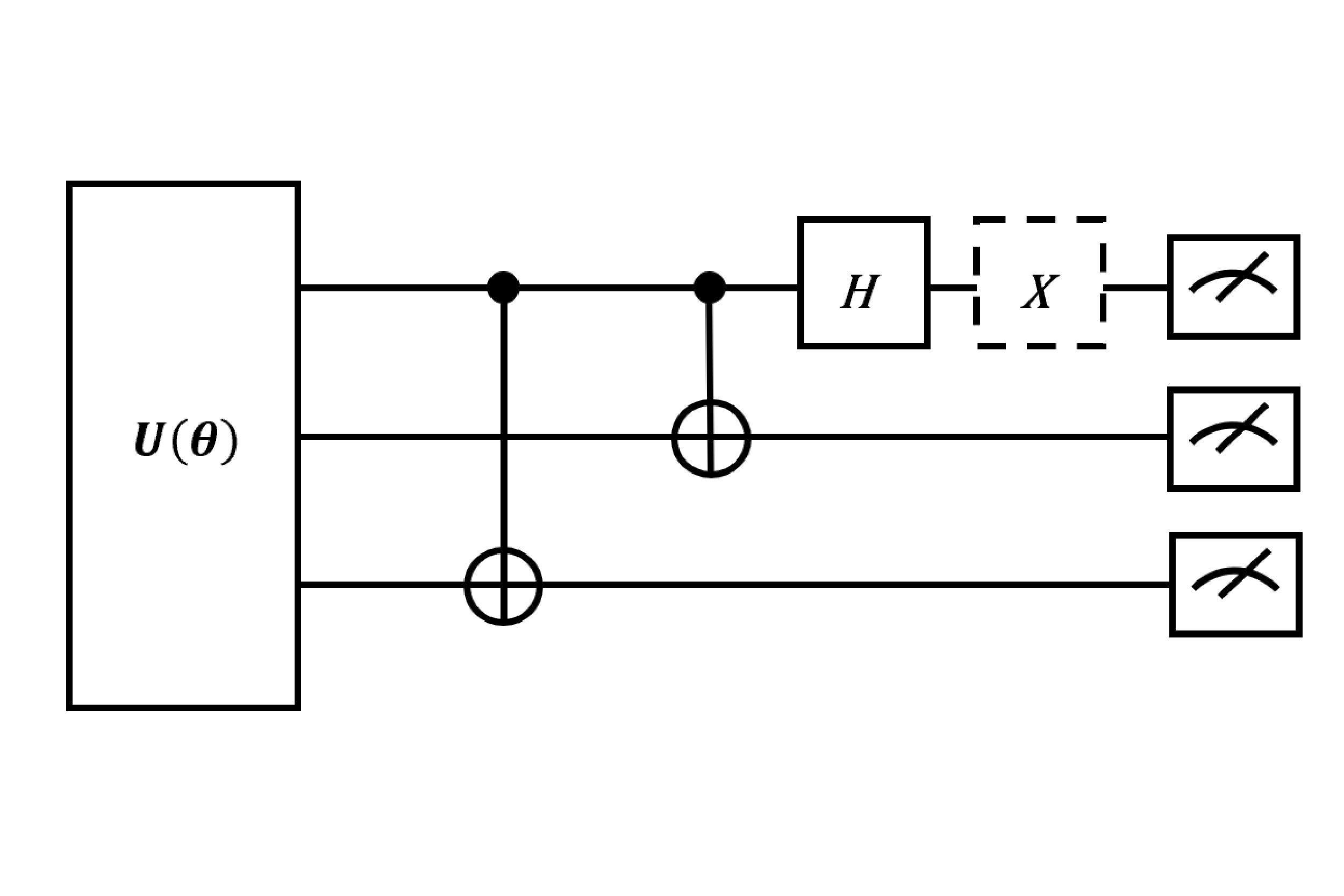}\label{fig1d}}
		\caption{The quantum circuits for estimating the values of cost function $E(\theta)$. (a) The Hadamard test circuit to estimate the items $\langle 0, \psi(\theta)|F_n^{-1}D^lF_n|0, \psi(\theta)\rangle$. (b) The Hadamard test circuit to estimate the items $\langle 0,b|F_n^{-1}D^lF_n|0,\psi(\theta)\rangle$. (c) Detailed structure of the unitary matrix $L^l$. (d) The quantum circuit of estimating $\langle\psi(\theta)|M_1|\psi(\theta)\rangle$.$\langle\psi(\theta)|M_1|\psi(\theta)\rangle$ can be evaluated by the probabilities of obtaining bit-strings 000 when measuring $H^{0}{\rm CNOT}_{1}^{0}{\rm CNOT}_{2}^{0}|\psi(\theta)\rangle$ and $X^{0}H^{0}{\rm CNOT}_{1}^{0}{\rm CNOT}_{2}^{0}|\psi(\theta)\rangle$ in the computational basis.}
\end{figure}

Similarly, we can obtain the decomposition of $\tilde{A}^2$ as
\begin{align*}
\tilde{A}^2=&\begin{bmatrix}
  6-4c-1+c^2&  -4+c&  1&  &  & 0 \\
  -4+c&  6&  -4&  1&  & \\
  1&  \ddots &  \ddots &  \ddots &\ddots   & \\
  &  \ddots &  \ddots &  \ddots &  \ddots & 1 \\
  &  &  1&  -4&  6& -4+d \\
  &  &  &  1&  -4+d& 6-4d-1+d^2
\end{bmatrix}\\
=&
\begin{bmatrix}
  6&  -4&  1&  &  & 0 \\
  -4&  6&  -4&  1&  & \\
  1&  \ddots &  \ddots &  \ddots &\ddots   & \\
  &  \ddots &  \ddots &  \ddots &  \ddots & 1 \\
  &  &  1&  -4&  6& -4 \\
  &  &  &  1&  -4& 6
\end{bmatrix}
-
\begin{bmatrix}
 4c+1-c^2&  &  &  &  &  \\
 &  0&  &  &  & \\
  &  & \ddots &   &  & \\
  &  &   &  \ddots &   &  \\
  &  &  &  &  0&  \\
  &  &  & & &4d+1-d^2
\end{bmatrix}\\
&+c\begin{bmatrix}
 0&  1&  0&  &  & 0 \\
 1&  0&  0&  &  & 0\\
  0&  \ddots &  \ddots &  \ddots &\ddots   & \\
  &  \ddots &  \ddots &  \ddots &  \ddots & 0 \\
  &  &  &  &  0&0 \\
  &  &  &  0& 0&0
\end{bmatrix}
+d\begin{bmatrix}
 0&  0&  0&  &  & 0 \\
 0&  0&  0&  &  & 0\\
  0&  \ddots &  \ddots &  \ddots &\ddots   & \\
  &  \ddots &  \ddots &  \ddots &  \ddots & 0 \\
  &  &  &  &  0& 1 \\
  &  &  &  0& 1&0
\end{bmatrix}\\
=&T_n^2-(4c+1-c^2)M_2-(4d+1-d^2)M_3+cM_4+dM_5.
\end{align*}

To summarize, for the one-dimensional Poisson equation with different boundary conditions, the number of decomposition terms of $\langle b|\tilde{A}|\psi(\theta)\rangle$ is $5$, and the number of decomposition terms of $\langle\psi(\theta)|\tilde{A}^2|\psi(\theta)\rangle$ is $6$. We provide the algorithm flowchart in Figure 2 for a more detailed presentation. The implementation of $T_n^1$ and $T_n^2$ are consistent with the previously mentioned approach, i.e., Eq.(\ref{eq1}) and Eq.(\ref{eq2}). Obviously, it is easy to obtain $\langle b|M_j|\psi(\theta)\rangle$, $j=\{1, 2\}$. As described earlier, $\langle\psi(\theta)|M_j|\psi(\theta)\rangle$, $j\in\{4,5\}$, can be obtained from the extended Bell measurements method. We take $n=8$ ($N=3$) for example to illustrate the specific quantum circuits too. The matrices can be rearranged as 
\begin{align*}
M_4&=|000\rangle\langle 001|+|001\rangle\langle 000|\\
&=\frac{1}{2} (|000\rangle+|001\rangle)(\langle 000|+\langle 001|)-\frac{1}{2 } (|000\rangle-|001\rangle)(\langle 000|-\langle 001|)\\
&=(H^{0}|000\rangle)(H^{0}|000\rangle)^\dagger-(H^{0}X^{0}|000\rangle)(H^{0}X^{0}|000\rangle)^\dagger,
\end{align*}
\begin{align*}
M_5&=|111\rangle\langle 110|+|110\rangle\langle 111|\\
&=\frac{1}{2} (|111\rangle+|110\rangle)(\langle 111|+\langle 110|)-\frac{1}{2} (|111\rangle-|110\rangle)(\langle 111|-\langle 110|)\\
&=(H^{0}|110\rangle)(H^{0}|110\rangle)^\dagger-(H^{0}X^{0}|110\rangle)(H^{0}X^{0}|110\rangle)^\dagger.
\end{align*}
And their expectation values can further be written as
\begin{align*}
\langle\psi(\theta)|M_4|\psi(\theta)\rangle
&=\langle\psi(\theta)|(H^{0}|000\rangle)(H^{0}|000\rangle)^\dagger-(H^{0}X^{0}|000\rangle)(H^{0}X^{0}|000\rangle)^\dagger|\psi(\theta)\rangle\\
&=|\langle\psi(\theta)|H^{0}|000\rangle|^2-|\langle\psi(\theta)|H^{0}X^{0}|000\rangle|^2,\\
\langle\psi(\theta)|M_5|\psi(\theta)\rangle
&=\langle\psi(\theta)|(H^{0}|110\rangle)(H^{0}|110\rangle)^\dagger+(H^{0}X^{0}|110\rangle)(H^{0}X^{0}|110\rangle)^\dagger|\psi(\theta)\rangle\\
&=|\langle\psi(\theta)|H^{0}|110\rangle|^2+|\langle\psi(\theta)|H^{0}X^{0}|110\rangle|^2.
\end{align*}

\begin{figure}[!h]
 \centering
 \includegraphics[width=1\linewidth]{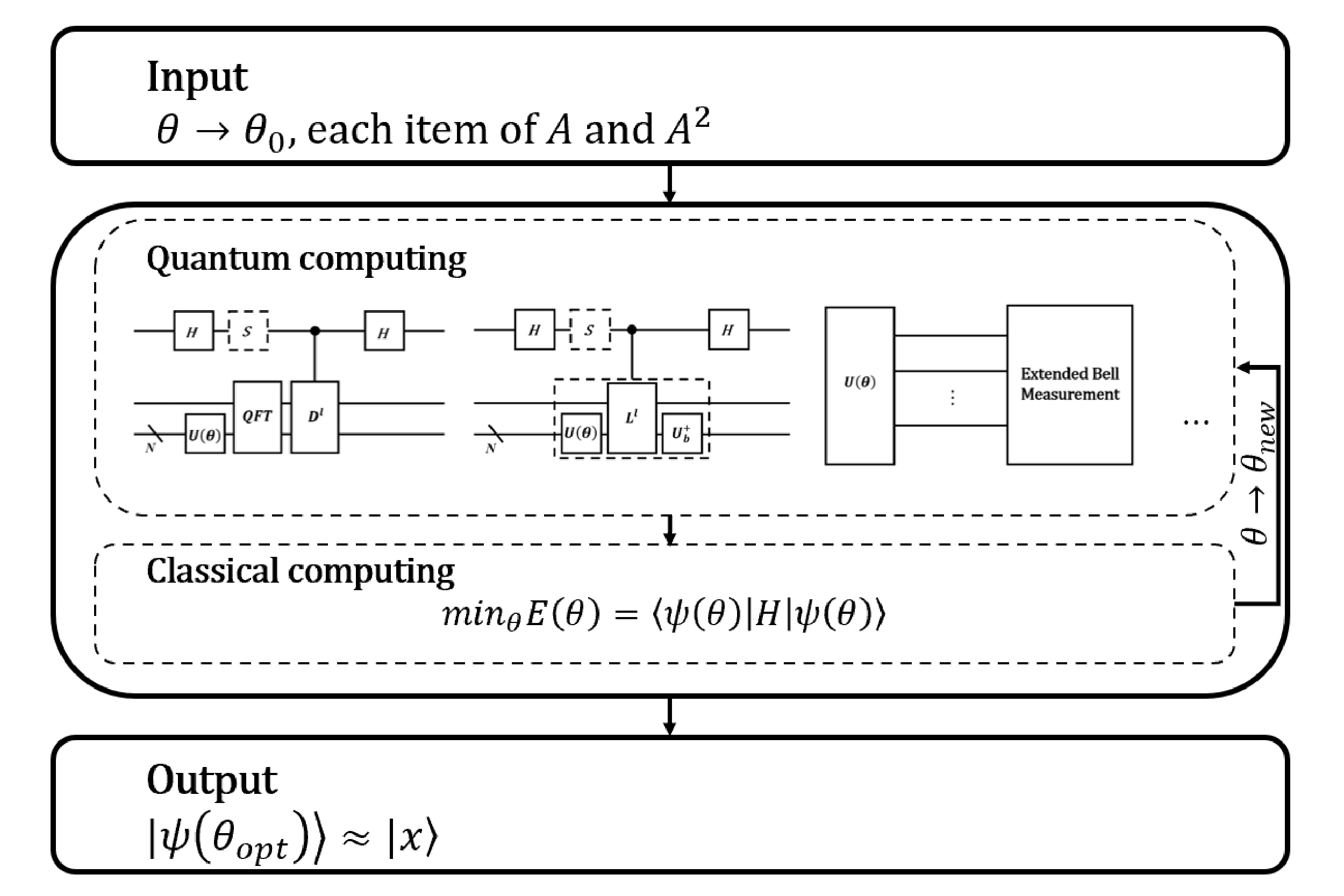}
 \caption{Schematic diagram of the entire algorithm for the one-dimensional Poisson equation with different boundary conditions.}
\end{figure}

\subsection{$d$-dimensional Poisson equation with Dirichlet boundary conditions}
Similarly, the $d$-dimensional Poisson equation with Dirichlet boundary conditions can be discretized to be the following linear system
$A^{(d)}x=b$,
where $b\in R^{n^d}$ and $A^{(d)}\in R^{n^d\times n^d}$,
with
\begin{align*}
A^{(d)}&=A'\otimes  I\otimes  \dots \otimes I+I\otimes A'\otimes I\otimes \dots \otimes I+\cdots +I\otimes \cdots \otimes I\otimes A'\\
&=\sum_{\substack{s=0\\s+t=d-1}}^{d-1} I_s\otimes A'\otimes I_t,
\end{align*}
where $I_s=\underbrace{I\otimes \dots \otimes I} _s$,  $I\in R^{n\times n}$.\\

Based on the structure of $A'$, it can be decomposed as,
\begin{gather*}
A'=\begin{bmatrix}
  2&  -1   &\cdots   &0  &0 \\
  -1&  2&  & \ddots    &0 \\
  \vdots & &\ddots   &  &\vdots  \\
  0&   &  &  \ddots &-1 \\
  0&0  &\cdots   &  -1&2
\end{bmatrix}
=2I-L-L^{-1}+M_5,
\end{gather*}
where
\begin{gather*}
M_5=\begin{bmatrix}
 &  &  &  &  &1  \\
 &  &  &  &  0& \\
 &  & \ddots &   &  & \\
 &  &   &  \ddots &   &  \\
 &  0&  &  &  &  \\
 1 &  &  & & &
\end{bmatrix}.
\end{gather*}\\

For matrix $M_5$ in $A^{(d)}$, we give the following decomposition, i.e.,
\begin{align*}
M_5&=\frac{1}{2}(\begin{bmatrix}
&  &  &  &  &1  \\
 &  &  &  &  1& \\
  &  & \ddots &   &  & \\
  &  &   &  \ddots &   &  \\
  &  1&  &  &  &  \\
 1 &  &  & & &
\end{bmatrix}
-\begin{bmatrix}
&  &  &  &  &-1  \\
 &  &  &  &  1& \\
  &  & \ddots &   &  & \\
  &  &   &  \ddots &   &  \\
  &  1&  &  &  &  \\
-1 &  &  & & &
\end{bmatrix})\\
&=\frac{1}{2}(
\begin{bmatrix}
&  &  &  &  &1  \\
 &  &  &  &  1& \\
  &  & \ddots &   &  & \\
  &  &   &  \ddots &   &  \\
  &  1&  &  &  &  \\
 1 &  &  & & &
\end{bmatrix}
-
\begin{bmatrix}
&  &  &  &  &1  \\
 &  &  &  &  1& \\
  &  & \ddots &   &  & \\
  &  &   &  \ddots &   &  \\
  &  1&  &  &  &  \\
1 &  &  & & &
\end{bmatrix}
\begin{bmatrix}
-1&  &  &  &  & \\
 & 1 &  &  &  & \\
  &  & \ddots &   &  & \\
  &  &   &  \ddots &   &  \\
  &  &  &  &1  &  \\
&  &  & & &-1
\end{bmatrix})\\
&=\frac{1}{2}\tilde{X}-\frac{1}{2}\tilde{X}\tilde{Z}.
\end{align*}\\

Since $M_5$ is contained in $A^{(d)}$, we will not continue the previous method of calculation. As mentioned in Theorem 2 of \cite{EBM}, for $s$-sparse matrices $X\in {\mathbb{C}}^{2^n\times 2^n}$, there exists such an algorithm that, in the worst case, the number of quantum circuits required to evaluate $\langle\psi| X |\psi \rangle$ is $2^{n+1}$. Thus, for $I_s\otimes M_5\otimes I_t$, the number of quantum circuits is $2^{Nd+1}$. Therefore, here we consider decomposing $M_5$ into a linear combination of two unitary matrices, i.e., $\tilde{X}$ and $\tilde{Z}$.\\

According to the above decomposition of matrix $A'$, we therefore obtain
\begin{align*}
A^{(d)}=&\sum_{\substack{s=0\\s+t=d-1}}^{d-1} I_s\otimes(2I-L-L^{-1}+\frac{1}{2}\tilde{X}-\frac{1}{2}\tilde{X}\tilde{Z})\otimes I_t\\
=&2dI^{\otimes d}+\sum_{\substack{s=0\\s+t=d-1}}^{d-1} I_s\otimes (-L-L^{-1}+\frac{1}{2}\tilde{X}-\frac{1}{2}\tilde{X}\tilde{Z})\otimes I_t,\\
\langle b|A^{(d)}|\psi(\theta)\rangle=&2d\langle b|\psi(\theta)\rangle+\sum_{\substack{s=0\\s+t=d-1}}^{d-1}(-\langle b|I_s\otimes L\otimes I_t|\psi(\theta)\rangle-\langle b|I_s\otimes L^{-1}\otimes I_t|\psi(\theta)\rangle\\
&+\frac{1}{2}\langle b|I_s\otimes \tilde{X}\otimes I_t|\psi(\theta)\rangle-\frac{1}{2}\langle b|I_s\otimes \tilde{X}\tilde{Z}\otimes I_t|\psi(\theta)\rangle).
\end{align*}\\

Similarly, we can obtain the decomposition of ${A^{(d)}}^2$ as
\begin{align*}
{A^{(d)}}^2=&2\sum_{\substack{u=0,v=0\\u+v+w=d-2}}^{d-2} I_u\otimes A'\otimes I_v\otimes A'\otimes I_w+\sum_{\substack{s=0\\s+t=d-1}}^{d-1} I_s\otimes A'^2\otimes I_t\\
=&2\sum_{\substack{u=0,v=0\\u+v+w=d-2}}^{d-2} I_u\otimes (2I-L-L^{-1}+\frac{1}{2}\tilde{X}-\frac{1}{2}\tilde{X}\tilde{Z})\otimes I_v\\
&\otimes (2I-L-L^{-1}+\frac{1}{2}\tilde{X}-\frac{1}{2}\tilde{X}\tilde{Z})\otimes I_w\\
&+\sum_{\substack{s=0\\s+t=d-1}}^{d-1} I_s\otimes (2I-L-L^{-1}+\frac{1}{2}\tilde{X}-\frac{1}{2}\tilde{X}\tilde{Z})^2\otimes I_t\\
=&(2\cdot \frac{d^2-d}{2}\cdot4I+2\sum_{\substack{u=0,v=0\\u+v+w=d-2\\U_1,U_2\in  \mathscr{M}_1}}^{d-2} I_u\otimes U_1\otimes I_v\otimes U_2\otimes I_w)\\
&+(\frac{25}{4}dI+\sum_{\substack{s=0\\s+t=d-1\\U_3\in  \mathscr{M}_2}}^{d-1} I_s\otimes U_3\otimes I_t)\\
=&(4d^2+\frac{9}{4}d)I+2\sum_{\substack{u=0,v=0\\u+v+w=d-2\\U_1,U_2\in  \mathscr{M}_1}}^{d-2} I_u\otimes U_1\otimes I_v\otimes U_2\otimes I_w\\
&+\sum_{\substack{s=0\\s+t=d-1\\U_3\in  \mathscr{M}_2}}^{d-1} I_s\otimes U_3\otimes I_t,\\
\langle\psi(\theta)|{A^{(d)}}^2|\psi(\theta)\rangle=&4d^2+\frac{9}{4}d\\
&+2\sum_{\substack{u=0,v=0\\u+v+w=d-2\\U_1,U_2\in  \mathscr{M}_1}}^{d-2}\langle\psi(\theta)|I_u\otimes U_1\otimes I_v\otimes U_2\otimes I_w|\psi(\theta)\rangle\\
&+\sum_{\substack{s=0\\s+t=d-1\\U_3\in  \mathscr{M}_2}}^{d-1}\langle\psi(\theta)|I_s\otimes U_3\otimes I_t|\psi(\theta)\rangle,
\end{align*}
where $\mathscr{M}_1=\{I,-L,-L^{-1},\frac{1}{2}\tilde{X},-\frac{1}{2}\tilde{X}\tilde{Z}\}$, and at most one of $U_1$ and $U_2$ is selected as $I\in R^{n\times n}$, $\mathscr{M}_2$ is the product of the elements of the set $\mathscr{M}_1$ and does not contain $I\in R^{n^d\times n^d}$. In fact, $\langle\psi(\theta)|I_s\otimes U_3\otimes I_t|\psi(\theta)\rangle$ only needs to compute 12 terms, i.e.,
\begin{align*}
&\langle\psi(\theta)|I_s\otimes U_3\otimes I_t|\psi(\theta)\rangle\\
=&\langle\psi(\theta)|I_s\otimes (-4L-4L^{-1}+2\tilde{X}-\frac{1}{2}\tilde{Z}+L^2+{L^{-1}}^2+\frac{1}{2}\tilde{Z}^2-2\tilde{X}\tilde{Z}-\frac{1}{2}L\tilde{X}-\frac{1}{2}\tilde{X}L^{-1}\\
&-\frac{1}{2}L^{-1}\tilde{X}-\frac{1}{2}\tilde{X}L+\frac{1}{2}L\tilde{X}\tilde{Z}+\frac{1}{2}\tilde{X}\tilde{Z}L^{-1}+\frac{1}{2}L^{-1}\tilde{X}\tilde{Z}+\frac{1}{2}\tilde{X}\tilde{Z}L)\otimes I_t|\psi(\theta)\rangle\\
=&-4(\langle\psi(\theta)|I_s\otimes L\otimes I_t|\psi(\theta)\rangle+\overline{\langle\psi(\theta)|I_s\otimes L\otimes I_t|\psi(\theta)\rangle}) \\
&+(\langle\psi(\theta)|I_s\otimes L^2\otimes I_t|\psi(\theta)\rangle+\overline{\langle\psi(\theta)|I_s\otimes L^2\otimes I_t|\psi(\theta)\rangle})\\
&-\frac{1}{2}(\langle\psi(\theta)|I_s\otimes L\tilde{X}\otimes I_t|\psi(\theta)\rangle+\overline{\langle\psi(\theta)|I_s\otimes L\tilde{X}\otimes I_t|\psi(\theta)\rangle})\\
&-\frac{1}{2}(\langle\psi(\theta)|I_s\otimes L^{-1}\tilde{X}\otimes I_t|\psi(\theta)\rangle+\overline{\langle\psi(\theta)|I_s\otimes L^{-1}\tilde{X}\otimes I_t|\psi(\theta)\rangle})\\
&+2\langle\psi(\theta)|I_s\otimes \tilde{X}\otimes I_t|\psi(\theta)\rangle-\frac{1}{2}(\langle\psi(\theta)|I_s\otimes \tilde{Z}\otimes I_t|\psi(\theta)\rangle+\frac{1}{4}\langle\psi(\theta)|I_s\otimes \tilde{Z}^2\otimes I_t|\psi(\theta)\rangle\\
&-2\langle\psi(\theta)|I_s\otimes \tilde{X}\tilde{Z}\otimes I_t|\psi(\theta)\rangle+\frac{1}{2}\langle\psi(\theta)|I_s\otimes L\tilde{X}\tilde{Z}\otimes I_t|\psi(\theta)\rangle+\frac{1}{2}\langle\psi(\theta)|I_s\otimes \tilde{X}\tilde{Z}L^{-1}\otimes I_t|\psi(\theta)\rangle\\
&+\frac{1}{2}\langle\psi(\theta)|I_s\otimes L^{-1}\tilde{X}\tilde{Z}\otimes I_t|\psi(\theta)\rangle+\frac{1}{2}\langle\psi(\theta)|I_s\otimes \tilde{X}\tilde{Z}L\otimes I_t|\psi(\theta)\rangle,
\end{align*}
where we use $L^\dagger=L^{-1}$ in the second equation.
To summarize, for the $d$-dimensional Poisson equation with Dirichlet boundary conditions, the number of decomposition terms of $\langle b|A^{(d)}|\psi(\theta)\rangle$ is $4d+1$, and the number of decomposition terms of $\langle\psi(\theta)|{A^{(d)}}^2|\psi(\theta)\rangle$ is $12d^2$. The computation of $\langle b|A^{(d)}|\psi(\theta)\rangle$ and $\langle\psi(\theta)|{A^{(d)}}^2|\psi(\theta)\rangle$ is shown in Figure 3. For the $d$-dimensional Poisson equation with the Dirichlet boundary conditions, as in \cite{poisson2023}, the number of terms is also independent of $n$, and the number of decompositions is the same for $\langle b|A^{(d)}|\psi(\theta)\rangle$. However, for the number of decompositions of $\langle\psi(\theta)|{A^{(d)}}^2|\psi(\theta)\rangle$, the number of terms is much less than $(4d+1)^2-(4d+1)$ in \cite{poisson2023}.
\begin{figure}[!h]
		\centering
        \subfigure[]{\includegraphics[width=0.5\textwidth]{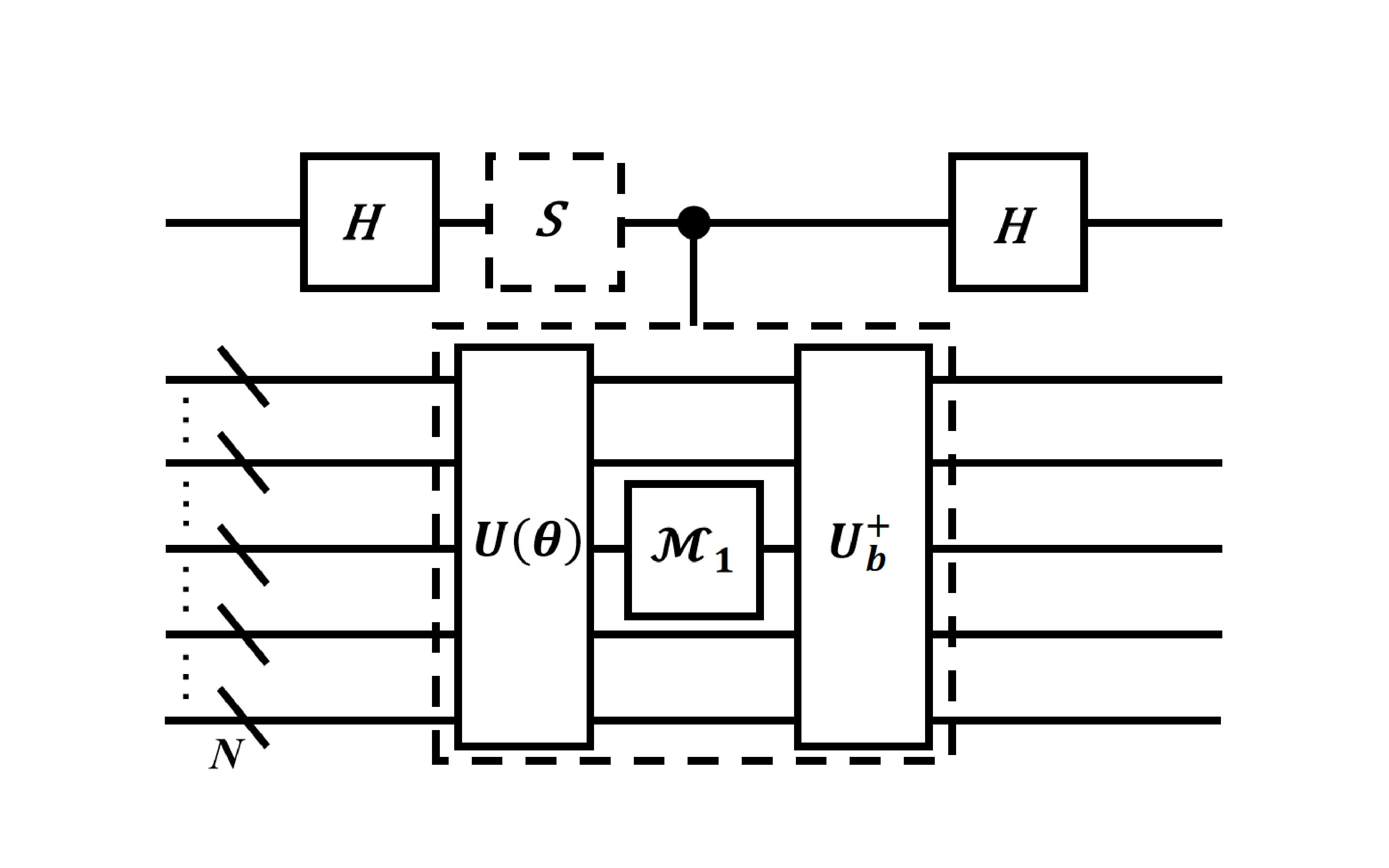}\label{fig3a}}\subfigure[]{\includegraphics[width=0.5\textwidth]{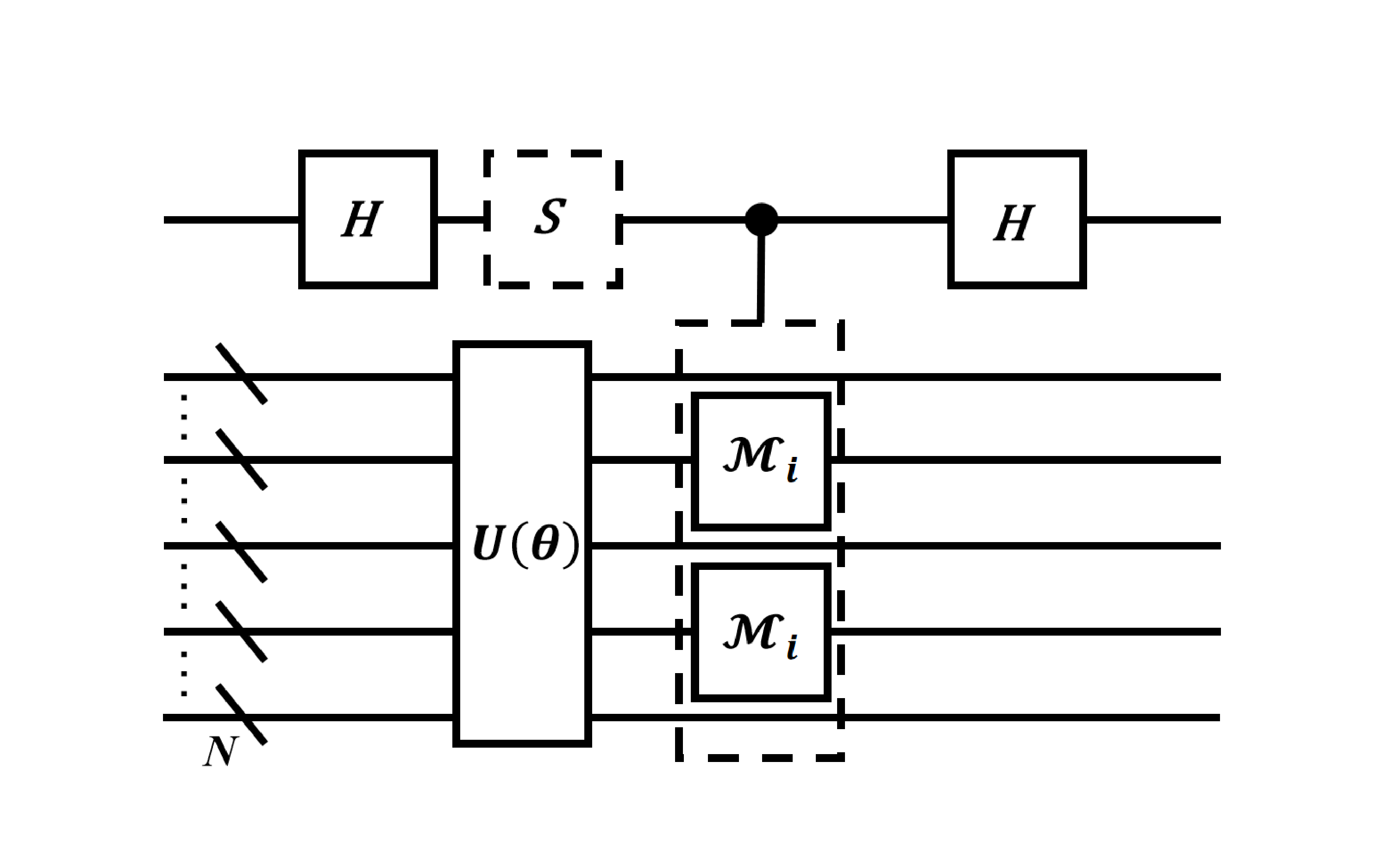}\label{fig3b}}
		\caption{The quantum circuits for estimating each term of $\langle b|A^{(d)}|\psi(\theta)\rangle$ and $\langle\psi(\theta)|{A^{(d)}}^2|\psi(\theta)\rangle$. (a)The Hadamard test circuit to estimate each term of $\langle b|A^{(d)}|\psi(\theta)\rangle$. (b) The Hadmard test circuit to estimate each term of $\langle\psi(\theta)|{A^{(d)}}^2|\psi(\theta)\rangle$. }
	\end{figure}

\subsection{Numerical simulation}
We perform numerical simulation in \textit{Python qiskit} package for the one-dimensional Poisson equation with Dirichlet boundary conditions to validate the feasibility for the proposed algorithms. By the finite-difference method, the discretized one-dimensional Poisson equation with Dirichlet boundary conditions is $A'x=b$. For this numerical simulation, we set $|b\rangle= {\textstyle \sum_{i=0}^{N}}(\frac{|0\rangle+|1\rangle}{\sqrt{2} } ) ^{\otimes N}$, which can be obtained by $U_b=H^{\otimes N}$, i.e., $|b\rangle=H^{\otimes N}|0\rangle$, where $N=3$. For ansatz, we chose the basic hardware-efficient ansatz in the numerical simulation shown in Figure 4. Since in our numerical simulation, our input data contains only real elements, and we choose $U(\theta_t^i)=R_y(\theta_t^i)$, as a way to reduce the training parameters. Otherwise, ansatz generally is choosed as $U(\theta)=R_z(\alpha _j )R_y(\beta _j )R_z(\gamma _j )$.\\

The main result is shown in Figure 5, which shows the depth of ansatz to be $2$. According to Figure 5, it can be seen that as the number of iteration step increases, the value of the cost function becomes smaller. The overall trend of the fidelity $ |\langle x |\psi(\theta)\rangle| $ tends to $1$ and the fidelity reaches more than 0.99. In addition, to avoid the cost function falling into local minimum, we solve this by choosing the parameters of $[0,2\pi]$ randomly several times.
\begin{figure}[!h]
 \centering
 \includegraphics[width=0.8\linewidth]{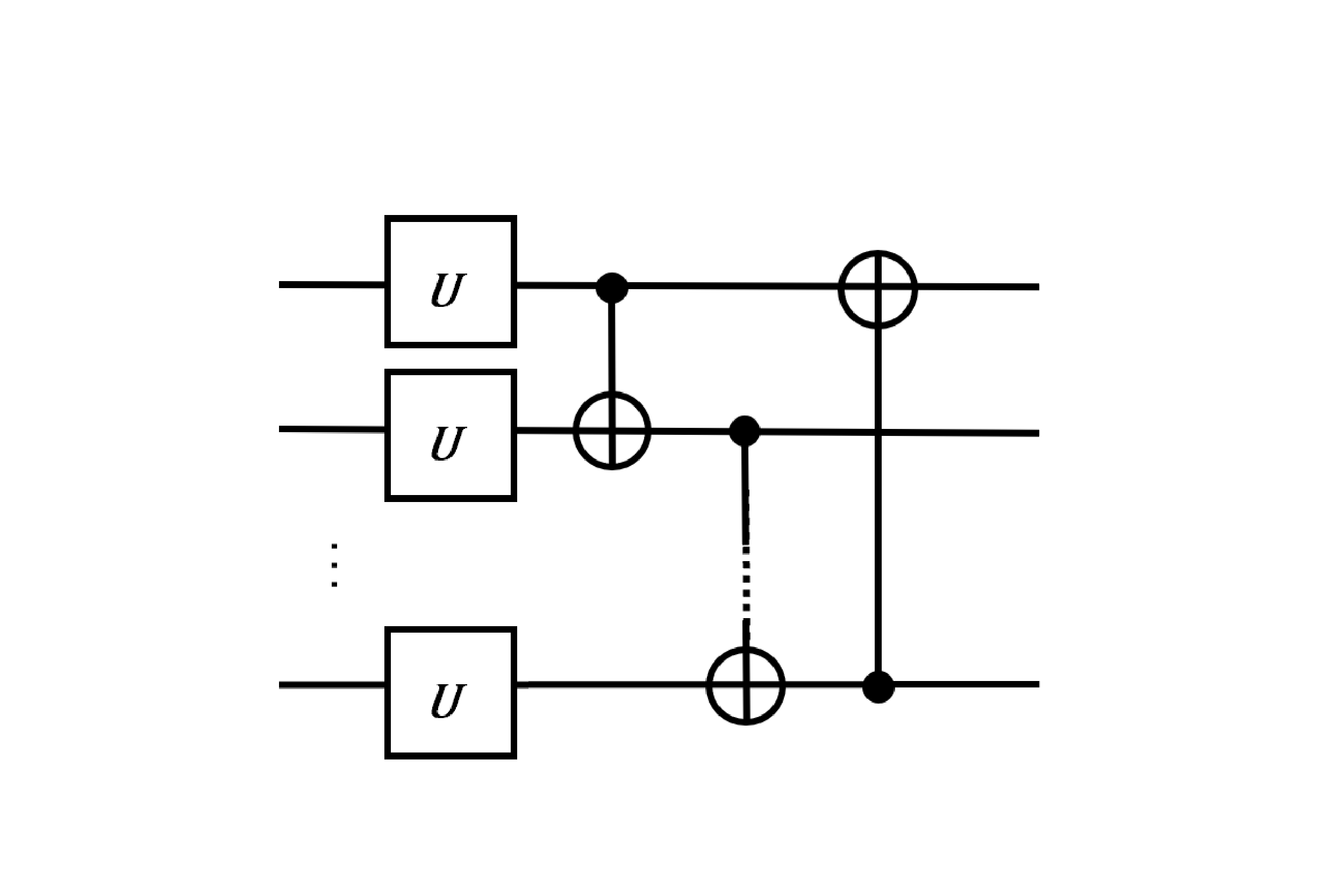}
 \caption{The basic hardware-efficient ansatz.}
\end{figure}
\begin{figure}[!h]
 \centering
 \includegraphics[width=0.8\linewidth]{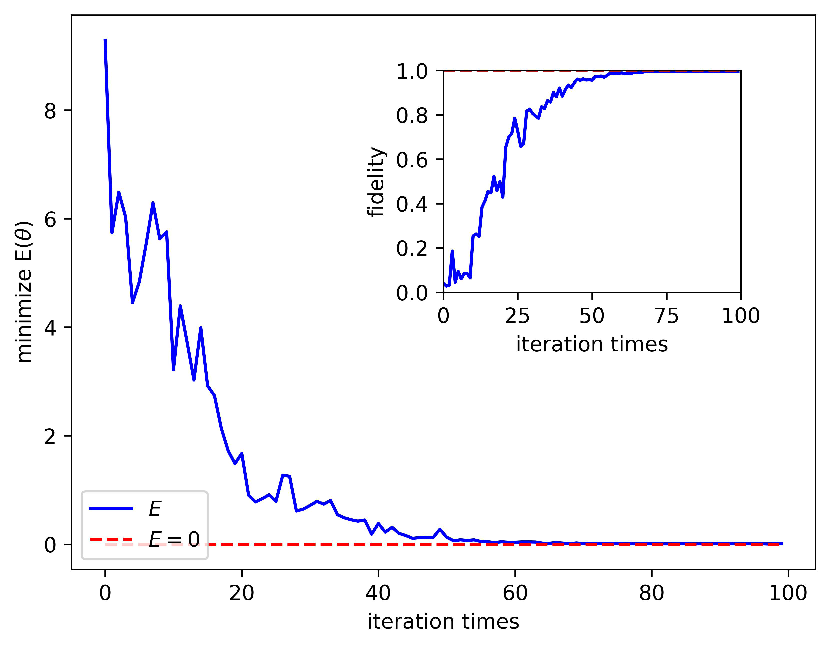}
 \caption{ The numerical simulation result of our algorithm. As the value of the cost function decreases, the value of fidelity $|\langle x |\psi(\theta)\rangle|$ converges to $1$.}
\end{figure}
\section{Variational algorithms for linear algebra operations with the banded Teoplitz matrix}
Gao et al. \cite{multiplication} have given an exact quantum algorithm on Toeplitz matrix-vector multiplication. In this section, we propose variational algorithms for linear algebra problems (including systems of linear equations and matrix-vector multiplication) with $K$-banded Teoplitz matrix.\\

For a given banded Teoplitz matrix $T_n^K$, $K\in O({\rm ploy}\log n)$, and an initial state vector $|v_0 \rangle$, we can compute the normalized state
\begin{gather*}
|v_T \rangle=\frac{T_n^K|v_0 \rangle}{\left \|T_n^K|v_0 \rangle  \right \| } ,
\end{gather*}
with $\left \|T_n^K|v_0 \rangle  \right \|=\langle v_0|T_n^{K\dagger}T_n^K|v_0\rangle$ and $T_n^K|v_0 \rangle \ne 0$. 
$|v_T \rangle$ can be found to be the corresponding eigenstate to the Hamiltonian 
\begin{gather*}
H_T=I-\frac{T_n^K|v_0 \rangle \langle v_0|T_n^{K\dagger}}{\left \|T_n^K|v_0 \rangle  \right \|^2},
\end{gather*}
with minimum eigenvalue $0$. And we define the cost function $ E(\theta)=\langle\psi(\theta)|H_T|\psi(\theta)\rangle$, that is, 

\begin{align*}
\underset{\theta}{\min} E(\theta)&=\underset{\theta}{\min}\langle\psi(\theta)|H_T|\psi(\theta)\rangle\\
&=\underset{\theta}{\min}\langle\psi(\theta)|(I-\frac{T_n^K|v_0 \rangle \langle v_0|T_n^{K\dagger}}{\left \|T_n^K|v_0 \rangle  \right \|^2})|\psi(\theta)\rangle\\          
&=1-\underset{\theta}{\min}\langle\psi(\theta)|\frac{T_n^K|v_0 \rangle \langle v_0|T_n^{K\dagger}}{\left \|T_n^K|v_0 \rangle  \right \|^2}|\psi(\theta)\rangle \\
&=1-\underset{\theta}{\min}\frac{|\langle\psi(\theta)|T_n^K|v_0 \rangle|^2}{\left \|T_n^K|v_0 \rangle  \right \|^2}\\
&=1-\underset{\theta}{\min}|\langle\psi(\theta)\frac{T_n^K}{\left \|T_n^K|v_0 \rangle  \right \|}|v_0 \rangle|^2\\   
&=1-\underset{\theta}{\min}|\langle\psi(\theta)\tilde{T_n^K} |v_0 \rangle|^2\\
&= 1-\underset{\theta}{\min}|\langle0,\psi(\theta)C_{\tilde{T_n^K}} |0,v_0 \rangle|^2,          
\end{align*}
where $\tilde{T_n^K}=\frac{T_n^K}{\left \|T_n^K|v_0 \rangle  \right \|}$ and 
\begin{gather*}
C_{\tilde{T_n^K}}=\begin{bmatrix}
  \tilde{T_n^K}&B\{\tilde{T_n^K}\}_n   \\
  B\{\tilde{T_n^K}\}_n&\tilde{T_n^K}
\end{bmatrix}.
\end{gather*}

The linear system with banded Teoplitz matrix $T_n^K$  is 
\begin{gather*}
T_n^Kx=b.
\end{gather*}
We also transform the problem of solving the linear system to find the ground state of a Hamiltonian,
$H={T_n^K}^\dagger (I-|b \rangle \langle b| )T_n^K $,
where $\left | b  \right \rangle \propto b$. And we assume that there is an efficient unitary operator $U_b$ that can prepare a quantum state $\left | b  \right \rangle$. Therefore, the cost function can be expressed as
\begin{align*}
		\underset{\theta}{\min} E(\theta)&=\underset{\theta}{\min}\langle\psi(\theta)|H|\psi(\theta)\rangle\\
                                           &=\underset{\theta}{\min}[\langle\psi(\theta)|{T_n^K}^2|\psi(\theta)\rangle-\left  |\langle b|T_n^K|\psi(\theta)\rangle\right | ^2]\\
&=\underset{\theta}{\min}[\langle0,\psi(\theta)|C_{{T_n^K}^2}|0,\psi(\theta)\rangle-\left  |\langle 0,b|C_{T_n^K}|0,\psi(\theta)\rangle\right | ^2],
\end{align*}
where
\begin{gather*}
C_{{T_n^K}^2}=\begin{bmatrix}
  {T_n^K}^2&B\{{T_n^K}^2\}_n   \\
  B\{{T_n^K}^2\}_n&{T_n^K}^2
\end{bmatrix},
\end{gather*}
and
\begin{gather*}
C_{T_n^K}=\begin{bmatrix}
  T_n^K&B\{T_n^K\}_n   \\
  B\{T_n^K\}_n&T_n^K
\end{bmatrix}.
\end{gather*}

The estimation of the cost function involved in this section can be obtained from the previous quantum circuits. Thus, equations such as the heat equation, Poisson equation and other equations involved in engineering, when discretised, can be transformed into linear system with the banded Teoplitz matrix solved and combined with the extended Bell measurements.
\section{Conclusion}
In this paper, we focus on solving the one-dimensional Poisson equation with different boundary conditions and the $d$-dimensional Poisson equation with Dirichlet boundary conditions on NISQ devices and further give the VQAs for linear algebra with the banded Teoplitz matrix. As in the case of general linear system solutions, we transform the solution of the discreted Poisson equation into a problem of solving its ground state in terms of Hamiltonian associated with the coefficient matrix $A$, and we use the structural features of  $A$ and $A^2$ to efficiently evaluate the value of the cost function. In detail, we use the structural features of the matrices $A$ and $A^2$ to decompose into a linear combination of the banded Teoplitz matrices and few sparse matrices. Then, for the one-dimensional Poisson equation with different boundary conditions and the $d$-dimensional Poisson equation with Dirichlet boundary conditions, the number of terms to decompose $\langle b|\tilde{A}|\psi(\theta)\rangle$ and $\langle b|A^{(d)}|\psi(\theta)\rangle$ is $5$ and $4d+1$ respectively, and the number of terms to decompose $\langle\psi(\theta)|\tilde{A}^2|\psi(\theta)\rangle$ and $\langle\psi(\theta)|{A^{(d)}}^2|\psi(\theta)\rangle$ is $6$ and $12d^2$ respectively. The quantum circuit consists of two main parts, one of which is the Hadamard test and the other is the extended Bell measurement. Using the Hadamard test to estimate the values (e.g. Figure 1(a), (b)) has a complexity of $O((N+1)^2)$ due to the QFT and the depth of QFT can also be reduced to $O(N)$ \cite{QFT}.  The quantum circuits require only 2 ancillary qubits, one of which is needed in order to extend the $A$ to the circulant matrix, and the other is needed for the Hadamard test. Moreover, for estimating $\langle 0, \psi(\theta)|F_n^{-1}D^lF_n|0,\psi(\theta)\rangle$, Figure 1(b) can also be solved using the generalized swap-test circuit as mentioned in \cite{GSWAP}. More importantly, our algorithm only requires changing the part of the circuit involving the parameter $D$ for estimating $\langle 0, \psi(\theta)|F_n^{-1}D^lF_n|0, \psi(\theta)\rangle$. The complexity of estimating $M_i$ using the extended Bell measurement method is at most $O(N)$ and does not require ancillary qubits. Finally, the quantum circuits have been explicitly constructed to evaluate efficiently the value of the cost function.\\

We emphasis that our algorithm for evaluating the cost function is very efficient since the number of decomposition terms is only polynomial of $d$ and independent of $n$. As a result, our algorithm greatly reduces the number of measurements compared to the algorithm in \cite{poisson2021}. Like the algorithm in \cite{poisson2023}, our approach does not depend on the size of the dimension $n$ of the linear system. Furthermore, we reduce the number of decomposition terms and provide the simpler circuit construction. In this paper, variational quantum algorithms for linear systems with Teoplitz matrices are proposed, which are adapted for Hermitian and sparse coefficient matrices satisfying $a_{i,i+c}=a_c$, for all $c=0,1,\cdots ,n-1$ and $i=0,\cdots ,n-1-c$, where $a_{i,i+c}$ denotes the element of the coefficient matrices mentioned in \cite{poisson2023}.\\

\section*{Acknowledgments}
This work is supported by the Shandong Provincial Natural Science Foundation for Quantum Science No. ZR2021LLZ002 and the Fundamental Research Funds for the Central Universities No. 22CX03005A.

\section*{\bf Data availability statement}
No data was used for the research described in the article.

\end{document}